\documentstyle[prb,aps,twocolumn,epsf]{revtex} 
\begin{document}
\title{Do we understand electron correlation effects in Gadolinium based intermetallic compounds?}

\author{E.V. Sampathkumaran and R. Mallik\footnote{Present address: Department of Developmental and Cell Biology, University of California, Irvine, USA.}}

\address{Tata Institute of Fundamental Research, Homi Bhabha Road, 
Colaba, Mumbai - 400005, India\\}

\maketitle

\begin{abstract} 
Recognising the difficulties in systematic understanding of the physical characteristics of strongly correlated f-electron systems, we considered it worthwhile to subject the so-called "normal" f-electron systems like those of Gd to careful investigations.  We find that the spin-disorder contribution to electrical resistivity ($\rho$) in the paramagnetic state, instead of remaining contant, surprisingly increases with decreasing temperature (T) in the paramagnetic state in some of the Gd alloys. In some cases, this "excess resistance" is so large that a distinct minimum in the plot of $\rho$ versus T can be seen, mimicking the behaviour of Kondo lattices. This excess resistance can be suppressed by the application of a magnetic field, naturally resulting in large magnetoresistance.  In addition, these alloys  are found to exhibit heavy-fermion-like heat-capacity behavior. These unusual findings imply hither-to-unexplored electron correlation effects even in Gd-based alloys. 

\end{abstract}

\footnote{$^*$Present address: Department of Developmental and Cell Biology, University of California, Irvine, USA.}
\maketitle
\vskip1cm
\section{Introduction}
The physical characteristics of the materials containing some of the f-elements
like Ce, Eu, Yb and U, arising from strong electron correlation effects, have been of great 
interest during last two decades in condensed matter physics and the identification of new materials containing these ions often resulted in discovery of novel phenomena. But, the fact remains that, inspite of intense theoretical and experimental activities, 
there are considerable difficulties in systematic understanding of these properties even at a qualitative level. However, it has been recognised that the observed  anomalies are in some way related to the tendency of the f-orbital to get delocalised, in other words, 
to the capability of these f-ions to exist in more than one valence state. Naturally, the f-ions with deeply localised f-orbital, like Gd, are expected to behave normally without any unusual electron correlation effects known for the above-mentioned class of f-ions and therefore careful investigations on Gd systems by and large do not exist in the literature.  During the course of our investigations  of strongly correlated electron systems (SCES), we came across some Gd alloys in which the observations are so puzzling that this fundamental assumption in this field is questionable. This article, surveying our experimental findings with few examples, raises a question whether the understanding of the electron correlation effects in "normal" rare-earths  is truly complete.


\section{Unusual results in ternary Gd compounds derived from AlB$_2$-derived  hexagonal structure}
A systematic investigation of Gd based alloys was actually motivated by our original observation of heavy-fermion-like heat-capacity (C) anomalies in Gd-substituted LaCu$_2$Si$_2$ \cite{EVS1995a}, similar to that known for Ce alloys in the low-coupling limit \cite{WPB1991}. Subsequently, we noticed that the electrical resistivity ($\rho$) of GdNi$_2$Si$_2$ \cite{EVS1995b} and GdNi \cite{Mallik1997} over a wide temperature (T) range above respective magnetic ordering temperatures (T$_o$) could be suppressed to relatively  smaller values by the application of a magnetic field (H). In other words, these Gd compounds are found to exhibit negative magnetoresistance (MR), the magnitude of which increases with decreasing temperature.   This finding implies that there is a T-dependent $\rho$-component in excess of lattice contribution present in these materials. While all these features, {\it new as far as Gd alloys are concerned}, are typical of Kondo alloys, no distinct minimum in the T dependence of $\rho$ could be observed in the raw experimental data. It is therefore of interest to look for Gd alloys which can directly show this minimum in the raw data. In this respect, the ternary Gd compounds, crystallizing in the AlB$_2$-derived hexagonal structure, provided an unique situation to demonstrate this aspect as well. While the reader may refer to all our published articles for various measurements  and discussions \cite{Mallik1998a,Mallik1998b,Mallik1998c,Saha1999,Subham1999a,Subham1999b,Subham1999c,Subham2000,Subham2001a,Chaika2001}, here we summarise the key experimental findings.      

\par
The behaviour of the compound Gd$_2$PdSi$_3$ \cite{Mallik1998a} by far remains as the best example to demonstrate our point.  We have confirmed by various measurements including $^{155}$Gd M\"ossbauer effect studies that the compound Gd$_2$PdSi$_3$ undergoes long range antiferromagnetic ordering at (T$_N$=) 21 K. As far as $\rho$(T) behaviour is concerned, what one would therefore expect is that the T-coefficient of $\rho$ is positive above 21 K. Looking at figure 1(a and c),  $\rho$(T) of Gd$_2$PdSi$_3$ gradually decreases as the T is lowered down to 60 K, however {\it followed by an upturn below about 45 K}, thus giving rise to a resistivity minimum at about (T$_{min}$) 45 K. The value of $\rho$ before the onset of magnetic ordering (say, at 22 K) is about 5\% higher compared to that at T$_{min}$.

Such a behaviour of $\rho$ with a distinct minimum in the paramagnetic state was not previously observed  for any "normal" (that means, with deeply localised f-orbital) f-ion based intermetallic compound. [Further lowering of T below T$_N$ does not result in a drop in $\rho$(T), as expected due to the loss of spin-disorder contribution, and this finding is not relevant to the present discussion; this is understandable in terms of opening of a pseudogap at some portions of the Fermi surface due to the onset of antiferromagnetic ordering]. Substitution of Y for Gd results in a depression of T$_N$ as expected, but the minimum in $\rho$(T) persists (see Fig. 1a).  The fact that the minimum is observable also in the sample with the highest Y content (x= 0.8) excludes any explanation for the observed $\rho$ minimum at 45 K in Gd$_2$PdSi$_3$ in terms of  the opening of an energy gap above 21 K. We have also estimated the 4f-contribution ($\rho_{4f}$) to $\rho$ by subtracting the $\rho$ of Y analogue; for this purpose, we adjusted the slope of $\rho$(T) plot with that of Gd$_2$PdSi$_3$ at 100 K and then matching the $\rho$ value at 100 K; though such an analysis to subtract out lattice contribution is not free from ambiguities, one gets an idea  about T-dependence of "excess $\rho$" ($\rho_{4f}$= $\rho$(T)-$\rho_{lattice}$) from the derived data (shown in Fig. 1b). Irrespective of the nature of reference for lattice contribution, the temperature coefficient of $\rho_{4f}$  is found to be distinctly negative in the paramagnetic state. 

\par
We have also measured $\rho$ in the presence of an externally applied H. For x= 0.0, it is found that MR, defined as [$\rho$(H)-$\rho$(0)]/$\rho$(0), is  about -10\% at 40 K in the presence of, say, H= 70 kOe, which is very large for a normal intermetallic compound in this temperature range \cite{Mallik1998a}. Interestingly, the minimum in $\rho$(T) vanishes in the presence of H (say at 50 kOe) and $\rho$ exhibits a positive T coefficient in the entire range (Fig. 1b). The   
magnitude of MR keeps increasing with decreasing T for a given H. These features are characteristics of Kondo alloys as well. It is to be noted \cite{Mallik1998a} that a large MR was not observed even for the analogous Kondo compound, Ce$_2$PdSi$_3$, which emphasizes the importance of the finding in the Gd compound. Another noteworthy finding is that the  T at which MR starts taking noticeably large values (with a negative sign) decreases with decreasing Gd concentration \cite{Mallik1998a}, following the trend in T$_N$. This implies that {\it the observed anomalies are magnetic in origin}.

\par
With respect to the C data (Fig. 1d), apart from the peaks attributable to magnetic ordering at T$_N$, the 4f contribution (C$_m$) to C exhibits a tail extending over a wide T range above T$_N$. For instance, for x= 0.8, for which T$_N$ is lower than 2 K, the tail in C$_m$(T) extends to T as high as 20 K, with the value of C/T as large as 1 J/mol K$^2$ at 2 K, comparable to the behaviour of prototype heavy-fermion system, CeCu$_2$Si$_2$.  

\par
It is clear from the above results that the compound, Gd$_2$PdSi$_3$, exhibits all the characteristics of (antiferromagnetically) ordering Kondo lattice systems in $\rho$, MR and C data, including a resistivity minimum. These features could be confirmed even in single crystals of this compound \cite{Saha1999}. In addition, the thermopower also exhibits a large value at 300 K, typical of Kondo systems \cite{Mallik1998b}. In order to explore whether these anomalies are due to an artifact of spin-glass freezing setting in around 45 K, we have probed frequency-dependence of ac susceptibility as well as possible irreversibilities in zero-field-cooled and field-cooled dc susceptibility behaviour carefully \cite{Mallik1998b, Subham1999a}. The results obtained, apart from establishing that the magnetic ordering below 21 K is of a long-range type,   rule out any type of magnetic ordering above 21 K. Therefore, a question arises whether this compound could be the first Gd-based Kondo-lattice. Our photoemission data on the other hand revealed \cite{Chaika2001} that there is no 4f-derived feature at the Fermi level (E$_F$) and that 4f-level is placed about 8 eV well below E$_F$, thereby conclusively establishing that this compound can not be categorised as a Kondo alloy. We will return to further discussion on this "magnetic precursor effect" (T-dependent "excess" $\rho$ and C$_m$ above T$_o$) in Section 4. 
\par
Before closing this section, it is worth stating that two more isostructural Gd compounds, viz., Gd$_2$PtSi$_3$ \cite{Subham2001a} and Gd$_2$CuGe$_3$ \cite{Subham2000}, have been subsequently identified to exhibit properties resembling those of Gd$_2$PdSi$_3$. This suggests that the anomalies discussed here must be more common among Gd intermetallics.            

\section{Behaviour of some other Gd compounds}

The observation of unusual magnetic precursor effects in the ternary Gd compounds mentioned above raises many questions: (i) Are these anomalies specific to this crystal structure? (ii) Is it something to do with the layered nature of Gd ions in this crystal structure? (iii) Is it due to polarization of the d-band of some of the transition metal ions, like Pt or Pd, which are prone for induction of moments by the large moment-carrying neighbouring ions (like Gd)? (iv) Is this magnetic precursor effect anything to do with the type of magnetic structure at lower temperatures?  (v) Is there relationship between C and $\rho$ anomalies? In order to tackle these issues, we have investigated a large number of Gd compounds \cite{Mallik1998c} with different crystal structures. The results clearly established that there is no straightforward relationship between the observation of the "excess $\rho$" on the one hand, and the crystal structure or the type of transition metal and s-p ions present in the compound, on the other.   It was also established that possible onset of magnetic correlations within a layer before long-range magnetic order sets in can not be offered as the sole reason for excess $\rho$, as many of the compounds with  Gd layers are not characterized by this "excess $\rho$". Also, Gd compounds exhibiting excess $\rho$ fall into the category of both ferromagnets and antiferromagnets and therefore the nature of the magnetic structure below T$_o$ is not the determining factor. 

A further careful look at the features above T$_o$ in the  $\rho$ and C data revealed \cite{Mallik1998c} that the Gd alloys can be classified into three categories: (i) Magnetic precursor effects appearing in both $\rho$ and C; (ii) "Excess $\rho$" and hence "MR anomaly" are negligible, but C tail persists over a wide temperature range; (iii) Both "excess $\rho$" and "excess C" are absent.  The behaviour of the second category of Gd compounds is demonstrated in figure 2 by showing the data for GdCu$_2$Ge$_2$, crystallizing in the layered ThCr$_2$Si$_2$-type crystal structure, for which T$_N$ is about 13 K. It is obvious from this figure that, apart from small magnitude, the sign of MR is not even negative (unlike first category of Gd alloys) in the entire T range. 
\par
      
\section{Discussion}

The central experimental observation is that, in some Gd alloys, the spin-disorder contribution to $\rho$ in the paramagnetic state apparently is not a constant value, but increases gradually with decreasing T as one approaches magnetic ordering temperature, resulting in Kondo-lattice-like characteristics. At this juncture, it may be mentioned that similar observations have been made also for Eu$_2$CuSi$_3$ \cite{Subham1999c} and Tb$_2$PdSi$_3$ \cite{Mallik1998d} and therefore these anomalies must be more common among other rare-earths as well with deeply localised f-orbitals. Since this feature does not arise from the Kondo effect, the origin of this unusual magnetic precursor effect remains a puzzle. One can not attribute it to Fisher-Langer type of critical spin-fluctuations at T$_o$ \cite{Ma1976, Fisher1968}, considering that the T-range over which these features  above T$_o$ appear is too large to be explained by this idea; in addition, the absence of a relationship between C and $\rho$ anomaly in second category of Gd compounds is convincing enough to rule out this possibility.  It therefore appears that the "excess $\rho$" in first category Gd  compounds arises from a hitherto unrecognised phenomenon in moment-carrying metals above T$_o$.  We speculate that the formation "magnetic polarons"  (localisation or trapping of the conduction electron cloud polarised by the s-f exchange interaction)  above T$_o$ could be the root-cause of these anomalies. The proposed magnetic localisation could be triggered by Gd4f short range magnetic order and consequent spatial magnetic fluctuations, just as crystallographic disorder reduces the mobility of charge carriers. We have observed \cite{Saha1999} distinct anomalies in Hall coefficient around 100 K in single crystals of Gd$_2$PdSi$_3$ and it is not clear at present whether this signals possible magnetic localisation effects. The application of a magnetic field suppresses the spin fluctuations or alters the localization length resulting in an enhancement of the mobility of the electron cloud. This explains the suppression of the $\rho$ minimum in the paramagnetic state in a magnetic field. Below T$_C$, in ferromagnetic Gd compounds like GdNi \cite{Mallik1997}, the polarons are anyway itinerant as they are coupled ferromagnetically with the Gd 4f moment; as a result polaronic contribution can be turned off below T$_C$. It must however be stated that the concept of magnetic polarons have been generally ignored in metals in the literature due to the fact  that large carrier density is expected to make these polarons delocalised. Possibly, in poor metals, one may have to still invoke this concept. If this proposal is confirmed in the compounds under discussion, one should explore various factors determining the presence or the absence of such effects. Some of the decisive factors could be the relative magnitudes of mean free path, localization length and short range correlation length, in addition to the strength of polarization of the conduction band and carrier density.  
\par
Another outcome of the studies on Gd alloys is the observation of negative MR even at temperatures far above T$_o$ (till 3T$_o$), peaking at T$_o$ in some Gd alloys (not presented here). These materials turned out to be one of the first few intermetallics exhibiting giant magnetoresistance behavior in the paramagnetic state at rather high temperatures. This finding established that neither double-exchange nor Jahn-Teller effects (proposed for by-now well-known "manganites") is crucial to observe GMR behavior. Similar conclusions  were drawn based on the observation of GMR behaviour in the pyrochlore, Tl$_2$Mn$_2$O$_7$ \cite{Shimakawa1996, MAS1996}, for which the concept of magnetic polarons has been discussed \cite{PM1998}. It is to be stressed that Gd$_2$PdSi$_3$ turned out to be the first {\it normal intermetallic} compound exhibiting  Tl$_2$Mn$_2$O$_7$-like $\rho$(T) behaviour.      

\section{Conclusion}

The discussion presented in this article suffciently demonstrates that, Gd intermetallics, which have been traditionally considered uninteresting from the magnetism point of view, exhibit novel behaviour in some compounds with rich features in their physical properties, even in the paramagnetic state. Our results on Gd alloys bring out for the first time that the understanding of electron correlation effects even in such  "normal" materials is far from complete. We hope that our data on several Gd compounds will motivate  new theoretical approaches in magnetism. 
Whatever be the microscopic origin of the anomalies, the results offer a message: One has to be a bit cautious while attributing  low-temperature upturns in $\rho$ and C in  alloys of Ce, U or Yb non-Fermi liquid and Kondo effects, since these features may also be as a precursor effect to the magnetic transitions occuring at further lower temperatures. It is of interest to explore, both theoretically and experimentally, whether magnetic polarons can give rise to non-Fermi liquid phenomena.


\begin{figure}
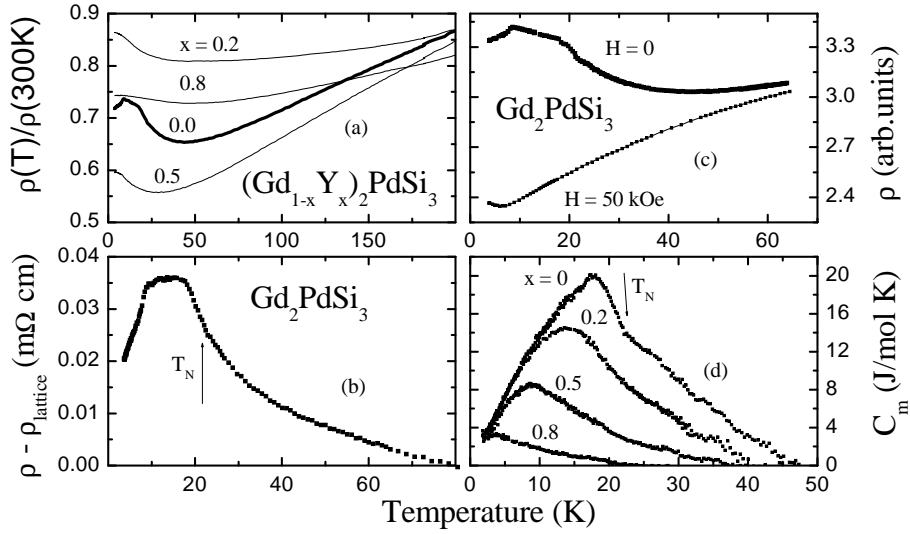

\caption{(a) The electrical resistivity ($\rho$) of the alloys, (Gd$_{1-x}$Y$_x$)$_2$PdSi$_3$, normalised to respective 300 K values, as a function of temperature (T). (b) An estimate of "excess $\rho$" as a function of T for Gd$_2$PdSi$_3$, obtained as described in the text. (c) The $\rho$ as a function of T for Gd$_2$PdSi$_3$ in the absence of a magnetic field and in the presence of 50 kOe. (d) The 4f-contribution to heat-capacity as a function of T for the alloys, (Gd$_{1-x}$Y$_x$)$_2$PdSi$_3$.} 
\end{figure}

\begin{figure}
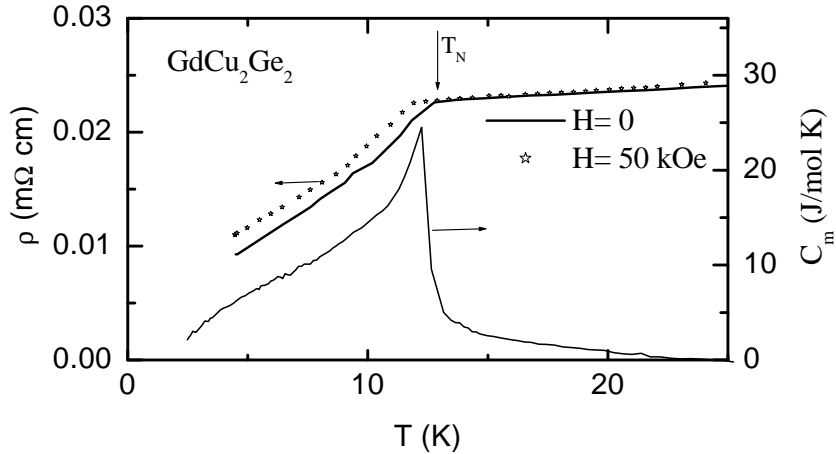

\caption{Electrical resistivity in zero magnetic field and in the presence of a magnetic field of 50 kOe and magnetic contribution (C$_{4f}$) to heat-capacity, for GdCu$_2$Ge$_2$, demonstrating that this compound belongs to Category-II as described in the text.}
\end{figure}

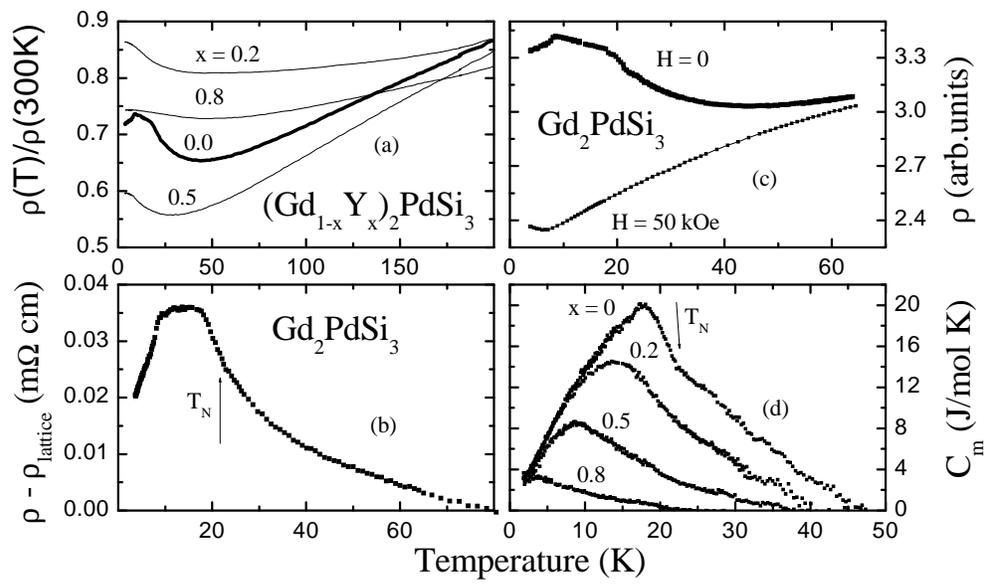
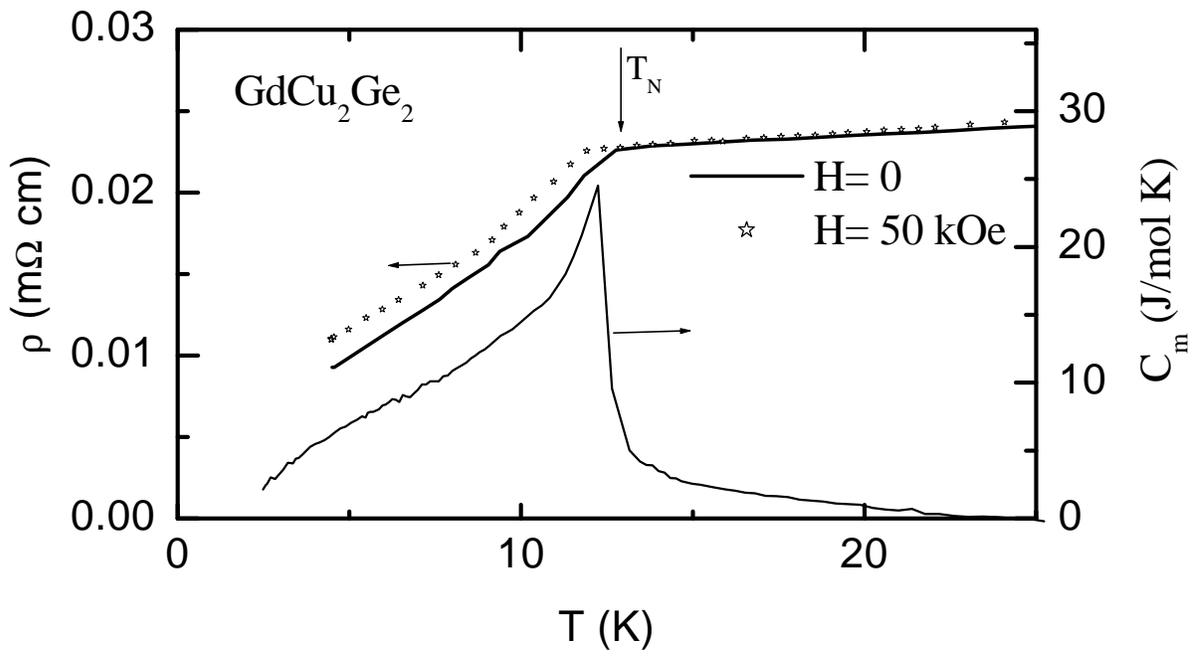
\end{document}